\documentclass[twocolumn,showpacs,preprintnumbers,amsmath,amssymb,aps,prb,superscriptaddress,longbibliography]{revtex4-2}
\usepackage{bbm}
\usepackage{mathrsfs}
\usepackage{graphicx}
\usepackage{dcolumn}
\usepackage{bm}
\usepackage{amsmath}
\usepackage{amsfonts}
\usepackage{color}
\usepackage[colorlinks=true,linkcolor=magenta,urlcolor=magenta,citecolor=cyan,anchorcolor=blue]{hyperref}
\usepackage{floatrow}

\begin{document}

\title{Isolated nearly flat higher Chern band in monolayer transition metal trihalides}
\author{Kejie Bao}
\affiliation{State Key Laboratory of Surface Physics and Department of Physics, Fudan University, Shanghai 200433, China}
\affiliation{Shanghai Research Center for Quantum Sciences, Shanghai 201315, China}
\author{Huan Wang}
\affiliation{State Key Laboratory of Surface Physics and Department of Physics, Fudan University, Shanghai 200433, China}
\affiliation{Shanghai Research Center for Quantum Sciences, Shanghai 201315, China}
\author{Jiaxuan Guo}
\affiliation{State Key Laboratory of Surface Physics and Department of Physics, Fudan University, Shanghai 200433, China}
\author{Yadong Jiang}
\affiliation{State Key Laboratory of Surface Physics and Department of Physics, Fudan University, Shanghai 200433, China}
\affiliation{Shanghai Research Center for Quantum Sciences, Shanghai 201315, China}
\author{Haosheng Xu}
\affiliation{State Key Laboratory of Surface Physics and Department of Physics, Fudan University, Shanghai 200433, China}
\affiliation{Shanghai Research Center for Quantum Sciences, Shanghai 201315, China}
\author{Jing Wang}
\thanks{Contact author: wjingphys@fudan.edu.cn}
\affiliation{State Key Laboratory of Surface Physics and Department of Physics, Fudan University, Shanghai 200433, China}
\affiliation{Shanghai Research Center for Quantum Sciences, Shanghai 201315, China}
\affiliation{Institute for Nanoelectronic Devices and Quantum Computing, Fudan University, Shanghai 200433, China}
\affiliation{Hefei National Laboratory, Hefei 230088, China}

\begin{abstract}
The interplay between non-trivial topology and strong electron interaction can generate a variety of exotic quantum matter. Here we theoretically propose that monolayer transition metal trihalides—WBr$_3$, WCl$_3$, WF$_3$ (with 2\% tensile strain) and MoF$_3$ (with 2\% tensile strain)—host isolated, nearly flat valence band near the Fermi level with a higher Chern number. The nontrivial topology of these flat Chern bands arises from an effective $sd^2$ hybridization of transition metal atom, which transform the apparent atomic $d$ orbitals on a hexagonal lattice into $(\tilde{s}, \tilde{p}_+, \tilde{p}_-)$ states on an emergent triangular lattice. Interestingly, the quantum geometry of flat Chern bands in these materials is comparable with that observed in moir\'e systems exhibiting fractional Chern insulator state. Furthermore, we investigate the Hofstadter butterfly spectrum of these flat Chern bands, revealing additional topological features. These natural materials hosting flat Chern bands with higher Chern numbers, if realized experimentally, could offer new playgrounds to explore exotic correlated phenomena.
\end{abstract}

\date{\today}

\maketitle

\section{Introduction}
Flat topological band systems have recently received significant attention as platforms to realize exotic correlated states. An outstanding example is fractional Chern insulator state, also called fractional quantum anomalous Hall state, was experimentally discovered in moir\'e materials recently in the absence of magnetic field~\cite{cai2023,zeng2023,park2023,xu2023,lu2024}. The prerequisites for fractional Chern insulator state is the presence of nearly flat and isolated Chern band with almost ideal quantum geometry~\cite{tang2011,sun2011,neupert2011,regnault2011,sheng2011,parameswaran2012,roy2014,claassen2015,ledwith2020,wang2021,parameswaran2013,ledwith2023,liu2024}. Flat topological bands support a lower bound on the superfluid density which otherwise would be zero in flat trivial bands~\cite{peotta2015}, and it was argued to participate in the superconductivity of moir\'e graphene~\cite{cao2018b,hu2019,julku2020,xie2020}. The experimental demonstration of flat Chern band (FCB) properties has been limited to only a few 2D moir\'e systems. As such, it is important and interesting to find stoichiometric 2D materials preferably in monolayer with isolated FCB near Fermi level, which could offer new and handier platforms to explore novel quantum states.

Searching for 2D materials that host isolated FCB is challenging because of the complex magnetic structures. It has been understood theoretically that FCB exists in line-graph of bipartite crystalline lattices with spin-orbit coupling (SOC)~\cite{mielke1991a,mielke1991b,tasaki1998,wu2007,bergman2008,liu2014,rhim2019,ma2020,calugaru2022,liu2021}. However, the predicted 2D materials with FCB are quite limited~\cite{liu2021,regnault2022,liu2013,yamada2016,sun2022,gao2023,pan2023,bhattacharya2023,neves2024,zhang2023,duan2024,ye2024,zhang2021bipolar,sheng2017monolayer,sun2019intrinsic,ouettar2022tuning}, and most of them share kagome geometry~\cite{yin2022} with Chern number of FCB to be $\mathcal{C}=1$, mimicking the lowest Landau level. The study of interaction effects in 2D kagome materials faces challenges, the principal of which being electron filling and stability of materials which favor the energetic position of flat band near Fermi level. Meanwhile, to explore new physics beyond lowest Landau level, it is interesting to have a FCB with higher Chern number~\cite{barkeshli2012,liu2012,wang2012,cooper2015,wang2022,ledwith2022,wang2011,yang2012,andrews2018,andrews2021,andrews2024}. There remains unclear whether FCB with higher Chern number can exist in realistic natural 2D materials, since generically long-range hopping are need to obtain higher Chern band with large flatness ratio.

\begin{table}[b]
\caption{Lattice constant, bandwidth ($W$), Chern number $\mathcal{C}$, fluctuation of Berry curvature $\delta\mathcal{B}$, and average trace condition violation $\mathrm{T}$ for the topmost valence band; Curie/N\'eel temperature $T_{c/n}$ from Monte Carlo simulations. $2\%$ tensile strain is applied to WF$_3$ and MoF$_3$.}
\begin{center}\label{tab1}
\renewcommand{\arraystretch}{1.5}
\begin{tabular*}{3.4in}
{@{\extracolsep{\fill}}ccccccc}
\hline
\hline
 &$a$ (\AA) & $W$ (meV) & $\mathcal{C}$ & $\delta\mathcal{B}$ & $\mathrm{T}$ & $T_{c/n}$ (K)\\
\hline
WBr$_3$ &6.64 & 51.5 & $-2$ & 0.89 & 0.15 & $T_n=60$
\\
WF$_3$  & 5.75 & 52.7 & $+2$ & 5.38 & 4.54 & $T_c=217$
\\
WCl$_3$ & 6.34 & 58.4 & $-2$ & 0.48 & 0.99 & $T_n=101$
\\
MoF$_3$ & 5.63 & 118.1 & $+3$ & 4.03 & 11.00 & $T_c=112$
\\
\hline
\hline
\end{tabular*}
\end{center}
\end{table}

Here we predict that monolayer WBr$_3$, WCl$_3$, WF$_3$ (with 2\% tensile strain), and MoF$_3$ (with 2\% tensile strain) have isolated flat band with higher Chern number near the Fermi level, based on density functional theory (DFT) calculations and the tight-binding model. These materials were identified through high-throughput screening in 2dMatPedia~\cite{zhou2019}. The Vienna \emph{ab initio} simulation package~\cite{kresse1996} is employed within Heyd-Scuseria-Ernzerhof (HSE) hybrid functional~\cite{krukau2006} to account for the correlation effect of $4d/5d$ electrons. Among these materials, WBr$_3$ and WF$_3$ serve as representative examples. WBr$_3$ exhibits an antiferromagnetic (AFM) ground state with a Ne\'el temperature of $60$~K, where the valence band forms a FCB with Chern number $\mathcal{C}=-2$ and a bandwidth of $51.5$~meV. Meanwhile, WF$_3$ under $2\%$ tensile strain stabilizes into a ferromagnetic (FM) ground state with a Curi\'e temperature of $217$~K, where the valence band also forms a FCB with Chern number $\mathcal{C}=+2$ and a bandwidth of $52.7$~meV. Interestingly, the topological nature of these flat bands arises from the band inversion within strongly hybridized cation $d$ orbitals on an effective triangular lattice.

\section{Structure and magnetic properties}
\begin{figure}[b]
\begin{center}
\includegraphics[width=3.4in, clip=true]{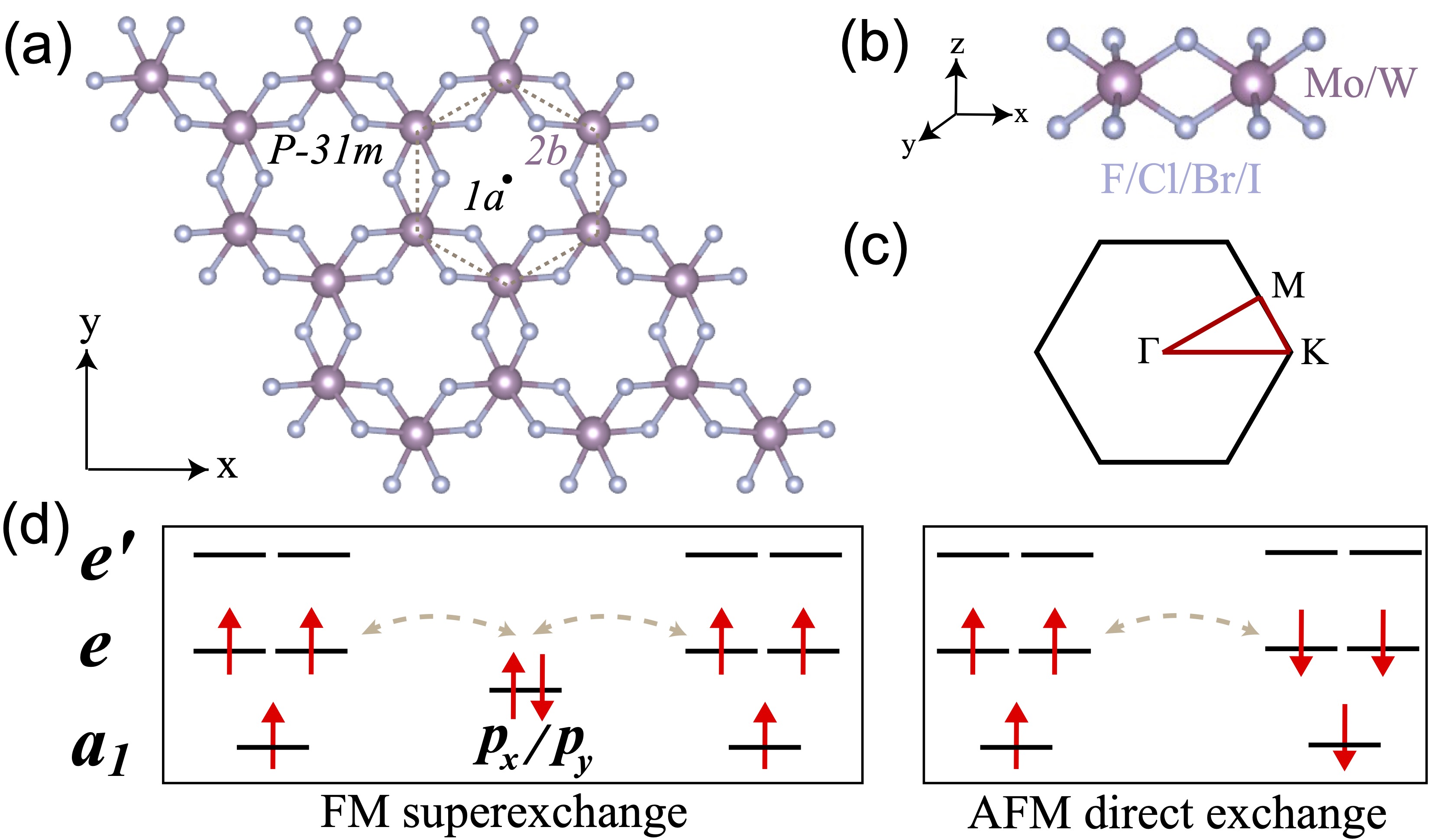}
\end{center}
\caption{(a),(b) Atomic structure of monolayer (Mo/W)$X_3$ ($X$=F, Cl, Br, I) from top and side views. The Wyckoff positions $1a$ and $2b$ are displayed  (notation adopted from Bilbao Crystallographic Server~\cite{bilbao2,bilbao3,slager2017,vergniory2017,elcoro2017,bradlyn2017}). The key symmetry operations of $P$-$31m$ include $C_{3z}$, $C_{2y}$ and inversion symmetry $\mathcal{I}$.  (c) Brillouin zone. (d) Crystal field splitting and schematic diagram of the direct exchange and superexchange between Mo/W $d$ electrons. }
\label{fig1}
\end{figure}

Monolayer transition metal trihalides crystallize in a honeycomb lattice with space group $P$-$31m$ (No.~162). As illustrated in Fig.~\ref{fig1}(a), using WBr$_3$ as a representative example, each primitive cell consists of three atomic layers, where each W atom is coordinated by six Br atoms, forming distorted edge-sharing octahedra. The lattice constants for these materials are summarized in  Table~\ref{tab1}. We focus mainly on WBr$_3$ and WF$_3$, while noting that similar results apply to other materials in this class~\cite{supple}. The dynamical and thermal stability of monolayer transition metal trihalides is confirmed by first-principles phonon and molecular dynamics calculations~\cite{supple}. Furthermore, the bulk phases of MoF$_3$ and MoCl$_3$ have been experimentally synthesized~\cite{lavalle1960,averdunk1990,hillebrecht1997structural}, suggesting a high likelihood of successfully fabricating their 2D counterparts. Notably, CrI$_3$ is particularly well-suited for the epitaxial growth of WBr$_3$ due to their structural similarity, where the lattice mismatch is only $\sim 3\%$ between CrI$_3$ (6.87~\AA) and WBr$_3$ (6.64~\AA)~\cite{li2020single}.

First-principles calculations reveal that WF$_3$ exhibits a strong FM ground state, while WBr$_3$ has a relatively weak Ne\'el AFM ground state. Both materials possess an out-of-plane magnetic easy axis.  The underlying mechanism of their magnetism can be understood from orbital occupation. These materials undergo a trigonal distortion, which involves compression along $z$ axis and elongation in $x$-$y$ plane. The distortion reduces the $O_h$ symmetry to $D_{3d}$ symmetry, resulting in the splitting of the W $5d$ orbitals into three levels: $a_1 (d_{z^2})$, $e$($d_{x^2-y^2}, d_{xy}$), and $e'$($d_{xz}, d_{yz}$) [Fig.~\ref{fig1}(d)]. The energy of $a_1$ and $e$ stays lower with respect to $e'$, as the latter point directly towards the negatively charged ligands. Consequently, each W atom adopts an $a_1^1e^2$ configuration, leading to a magnetic moment of $3\mu_B$ per W atom according to Hund's rule, consistent with DFT calculations. In these systems, two competing magnetic interactions determine the ground state. The nearly $90^\circ$ W–F–W and W–Br–W bonds favor FM superexchange~\cite{khomskii2004}, while direct exchange between W atoms promotes AFM. In WBr$_3$, the shorter W–W distance strengthens the AFM direct exchange, making it dominant and leading to AFM ground state with a N\'eel temperature of $60$~K. In contrast, in WF$_3$, FM superexchange prevails, leading to FM ground state with a Curi\'e temperature of $217$~K~\cite{supple}. Notably, in WBr$_3$, an external magnetic field can drive the AFM state into an FM state, where a FCB with a higher Chern number emerges.

\section{Electronic structures}
\begin{figure}[t]
\begin{center}
\includegraphics[width=3.4in, clip=true]{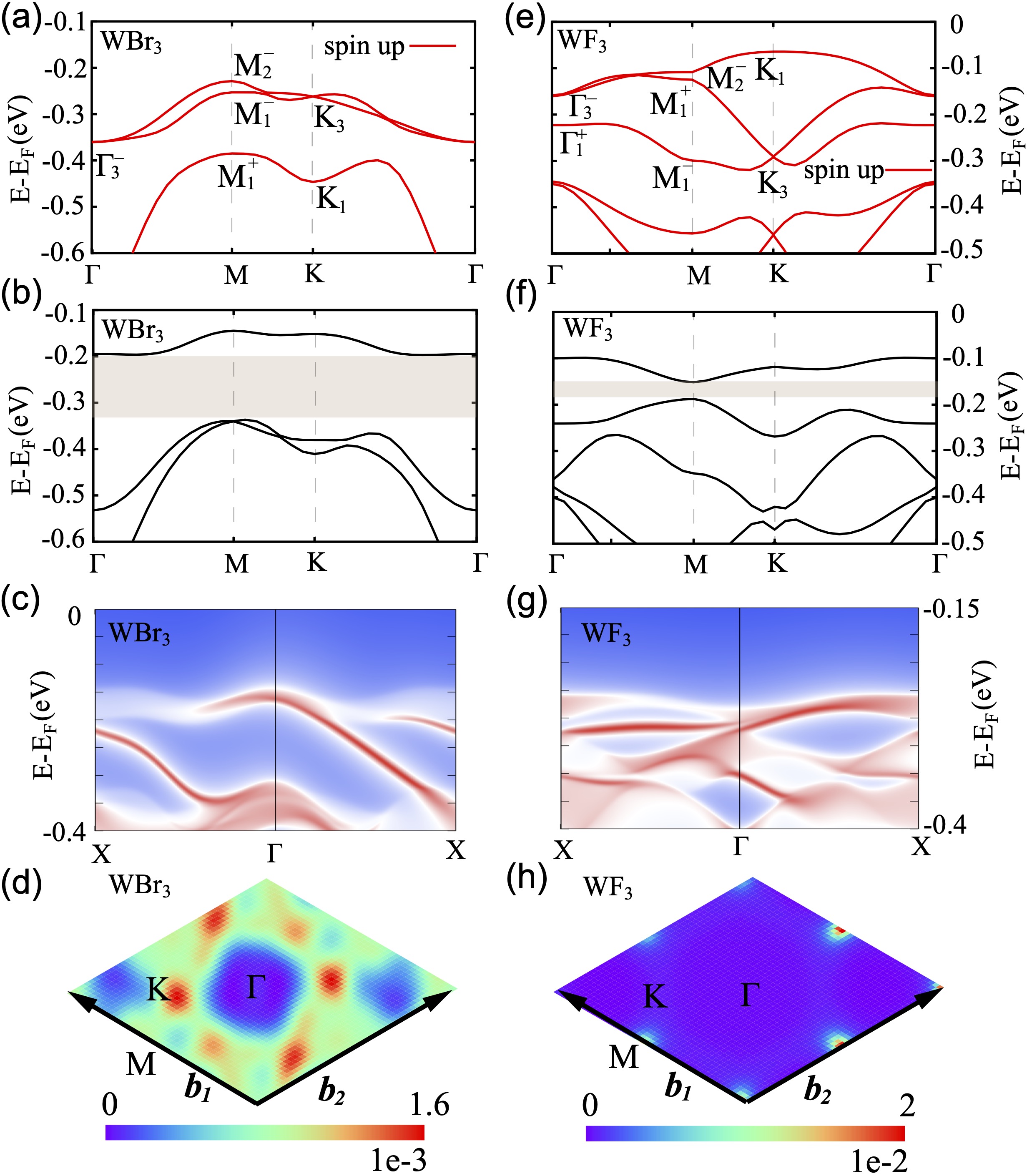}
\end{center}
\caption{Electronic structure and topological properties of monolayer WBr$_3$ and WF$_3$. (a-d) FM WBr$_3$, (e-h) FM WF$_3$ under $2\%$ tensile strain. The band structure without and with SOC, where the irreducible representation at high symmetry points are shown; topological edge states calculated along $x$-axis; Berry curvature of the first valence band, which exhibits a nearly FCB. The Berry curvature remains the same sign throughout the whole Brillouin zone in both WBr$_3$ and WF$_3$.}
\label{fig2}
\end{figure}

Fig.~\ref{fig2}(a) and Fig.~\ref{fig2}(b) display the electronic structure of monolayer WBr$_3$ in the FM state, without and with SOC, respectively. Near the Fermi level, three spin-up bands predominantly originate from W $d$ orbitals corresponding to the $a_1$ and $e$ levels. Band representation analysis reveals that these bands can be effectively described using the $(\tilde{s}, \tilde{p}_+, \tilde{p}_-)$ basis at the Wyckoff position $1a$ of an emergent triangular lattice, as summarized in Table~\ref{tab2}~\cite{supple}. The valence band edge exhibits $\tilde{p}_\pm$ character, featuring double degeneracies at $\Gamma$ and $K$, which are lifted by SOC. While SOC-induced gap openings in
$(\tilde{p}_+, \tilde{p}_-)$ on a triangular lattice typically lead to trivial topology~\cite{bradlyn2017}, Berry curvature calculations confirm a Chern number of $\mathcal{C}=-2$ of the valence band. This result is consistent with the presence of two chiral edge states within the gap, as depicted in Fig.~\ref{fig2}(c). Notably, this topological Chern band is nearly flat, with a bandwidth of about $W\approx51.5$~meV, and remains well isolated from other bands.

\begin{table}[b]
\caption{The elementary band representations of space group $P$-$31m$ for $(\tilde{s}, \tilde{p}_+, \tilde{p}_-)$ orbitals at Wyckoff positions $1a$.} 
\begin{center}\label{tab2}
\renewcommand{\arraystretch}{1.5}
\begin{tabular*}{3.4in}
{@{\extracolsep{\fill}}cccc}
\hline
\hline
   & $\Gamma$ & $M$ & $K$\\ 
\hline
   $\tilde{s}$ & $\Gamma_1^+(1)$ & $M_1^+(1)$& $K_1(1)$ \\
   $(\tilde{p}_+,\tilde{p}_-)$& $\Gamma_3^-(2)$ & $M_1^-(1)$ $\oplus$  $M_2^-(1)$& $K_3(2)$ \\
\hline
\end{tabular*}
\end{center}
\end{table}

Next, we examine WF$_3$. Fig.~\ref{fig2}(e) and Fig.~\ref{fig2}(f) show the electronic structure of the monolayer WF$_3$ in the FM ground state without and with SOC, respectively. A $2\%$ tensile strain is applied to 
enhance the isolation of flat bands~\cite{supple}. Similar to WBr$_3$, three spin-up bands near the Fermi level originate from W $d$ orbitals and are well described by the $(\tilde{s}, \tilde{p}_+, \tilde{p}_-)$ basis in a triangular lattice. However, unlike WBr$_3$, WF$_3$ exhibits six Dirac points along $\Gamma$-$M$ and double degeneracy at $\Gamma$ point, which are further gapped by SOC—indicating a nontrivial topology of the valence band. Berry curvature calculations and edge state analysis [Fig.~\ref{fig2}(g,h)] confirm a Chern number of $\mathcal{C}=+2$ for the valence band. This topological Chern band remains well isolated, with a small bandwidth of $W\approx52.7$~meV. The topological properties of WCl$_3$ and MoF$_3$ are further detailed in~\cite{supple}.

\section{Tight-binding model and analysis of topology}
\begin{figure}[b]
\begin{center}
\includegraphics[width=3.4in,clip=true]{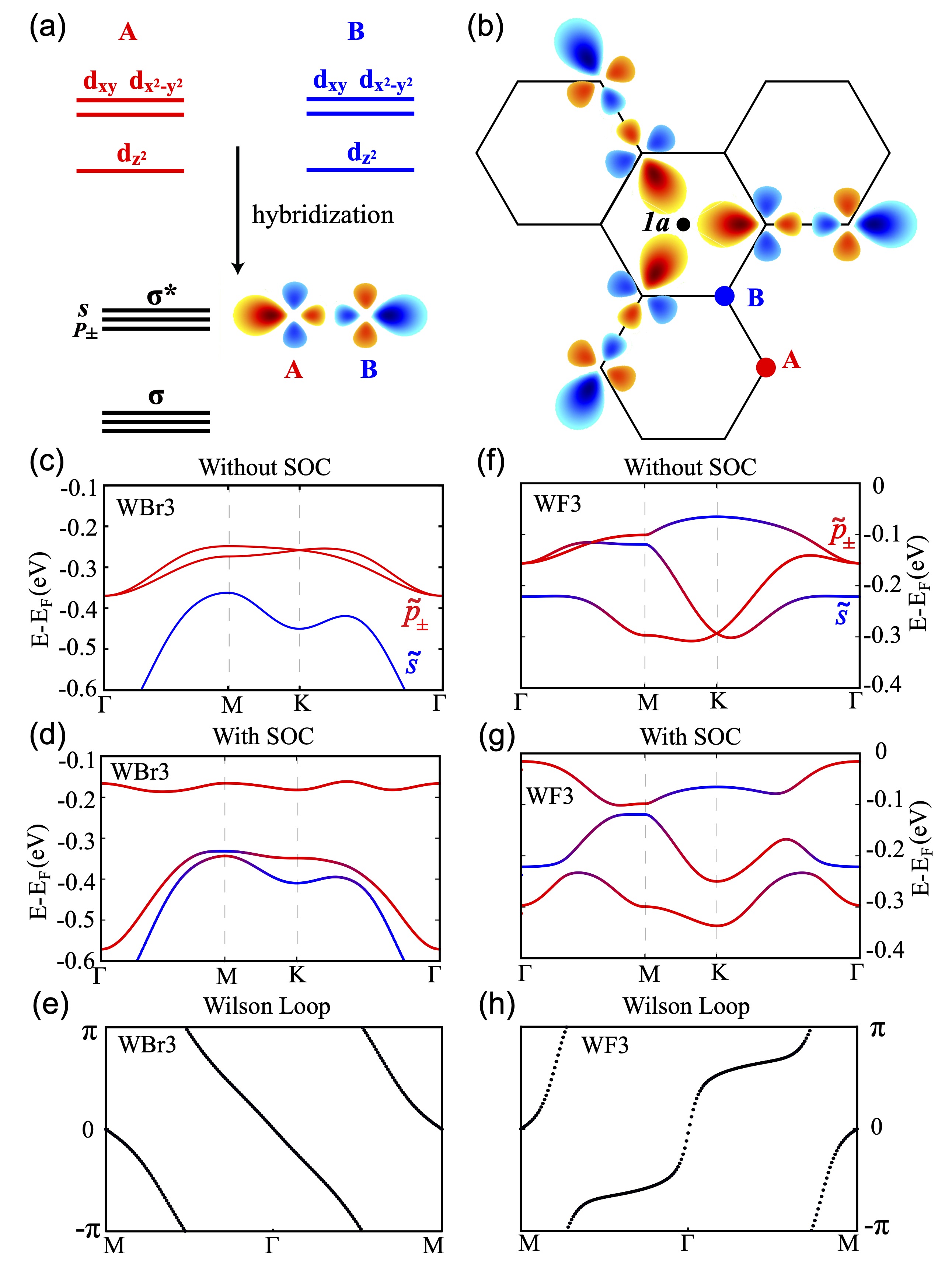}
\end{center}
\caption{(a) Schematics of $sd^2$ hybrid orbitals and antibonding $\sigma^*$ state. (b) Trianglular lattice formed by $\sigma^*$ state at Wyckoff position $1a$ as basis in a hexagonal lattice. (c-e) WBr$_3$, (f-h) WF$_3$. The band structure of tight-binding model with and without SOC, and Wilson loop for the topmost valence band. The irreducible representation at high-symmetry points of Brillouin zone boundary in (c) and (f) are consistent with that in Fig.~\ref{fig3}(a) and Fig.~\ref{fig3}(e), respectively. $~\tilde{s}$ and $\tilde{p}$ type bands are labeled as blue and red, respectively.}
\label{fig3}
\end{figure}

To uncover the physical origin of such peculiar electronic structures, we first conduct a detailed orbital analysis to demonstrate how the apparent atomic $d$ orbitals on a hexagonal lattice effectively transform into $(\tilde{s},\tilde{p}_+,\tilde{p}_-)$-type orbitals on a triangular lattice. The $d$-orbital projected band structure~\cite{supple} reveals that the natural orbitals $d_{z^2}$ (which appears $s$-like in planar view) and $(d_{xy},d_{x^2-y^2})$ with in-plane characteristics hybridize to form the bands of interest near the Fermi level. This process resembles $sd^2$ hybridization, which forms bonding $\sigma$ and antibonding $\sigma^*$ states with trigonal planar geometry [see Fig.~\ref{fig3}(a,b)], having the main contribution towards Wyckoff position $1a$. Interestingly, these hybrid states effectively transform the hexagonal symmetry of the atomic Mo/W lattice into the physics of a triangular lattice. Elementary band representation analysis further confirms this transformation: the states 
$(d_{z^2},d_{xy},d_{x^2-y^2})$ at Wyckoff position $2b$ map onto $(\tilde{s},\tilde{p}_+,\tilde{p}_-)^\pm$ at $1a$, thereby forming triangular lattice. The remaining three $d$-bands, also exhibiting $(\tilde{s},\tilde{p}_+,\tilde{p}_-)$-like characteristics, lie below the bands of interest~\cite{supple}.

To capture the essential topological physics, we construct a three-orbital tight-binding model that applies to both WBr$_3$ and WF$_3$, with material-specific parameters. This model is based on the $(\tilde{s}^\uparrow, \tilde{p}_+^\uparrow, \tilde{p}_-^\uparrow)$ orbitals located at Wyckoff position $1a$, as identified in the band representation analysis in Fig.~\ref{fig2}. The Hamiltonian is derived by incorporating nearest-neighbor and next-nearest-neighbor hopping terms, formulated using the Slater-Koster method, with SOC included~\cite{supple}. As demonstrated in Fig.~\ref{fig3}, this model successfully reconstructs the band structure and accurately reproduces the irreducible representations at high-symmetry points obtained from DFT calculations.

The different topology of FCB in monolayer WBr$_3$ and WF$_3$ originates from SOC, which can be explicitly written as 
\begin{eqnarray}\label{soc}
\mathcal{H}_{\text{soc}} &=& \lambda_{1}\sum_{i} (b_{i}^\dagger b_{i} - c_{i}^\dagger c_{i})+\lambda_{2}\sum_{\langle ij\rangle}[(b_{j}^\dagger b_{i} - c_{j}^\dagger c_{i}) + \mathrm{H.c.}]
\nonumber
\\
&&+\lambda_{3}\sum_{\langle ij\rangle}[(e^{-i\theta_{ij}}b_{j}^\dagger a_{i} - e^{i\theta_{ij}}c_{j}^\dagger a_{i}) + \mathrm{H.c.}],
\end{eqnarray}
where $(a_i, b_i, c_i)$ represent annihilation operators at site $i$ for $(\tilde{s}, \tilde{p}_+,\tilde{p}_-)$ orbitals, $\lambda_{1,2,3}$ are parameters from atomic SOC term as $\lambda_{\text{so}}\boldsymbol{\ell}\cdot\boldsymbol{\sigma}=\lambda_{\text{so}}(\ell_+\sigma_-+\ell_-\sigma_+)/2+\lambda_{\text{so}}\ell_z\sigma_z$. $\lambda_1=\lambda_{\text{so}}$ originates from the last term, and $\lambda_{2,3}$ arise from the first term, combined with orbital hopping and hybridization, $\lambda_{2,3}\sim\lambda_{\text{so}}^2/\Delta E$, meaning they appear as second-order processes with $\Delta E$ ($\sim1$~eV) being the energy difference between intermediate states and $(\tilde{s},\tilde{p}_\pm)$ orbitals. $\langle ij\rangle$ denotes the nearest-neighbor sites, $\theta_{ij}$ is the angle between $x$ axis and direction from site $i$ to $j$.  For WF$_3$, $\mathcal{H}_{\text{soc}}$ induces gap opening at six Dirac points along $\Gamma$-$M$ and quadratic band touching $\Gamma$ point. The six Dirac points, related by $C_{3z}$ and $\mathcal{I}$ symmetries, contribute Chern number $\mathcal{C}=3$. The $\Gamma$ point, on the other hand, contributes Chern number $\mathcal{C}=-1$. Thus, the total Chern number for FCB in WF$_3$ is $\mathcal{C}=3-1 = 2$, which is consistent with the Wilson loop calculation shown in Fig.~\ref{fig3}(h). For WBr$_3$, the $\lambda_1$ and $\lambda_2$ terms in Eq.~(\ref{soc}) induce gap openings at the quadratic band touching point $\Gamma$ and Dirac point $K$, where $\Delta_{\Gamma}=2\lambda_{\text{so}}+12\lambda_2$ and $\Delta_{K}=2\lambda_{\text{so}}-4\lambda_2$. Due to $\mathcal{I}$ symmetry, the Chern number of the valence band must be either: $\mathcal{C}=-2$ if $\text{sgn}(\Delta_{\Gamma}\Delta_K)=-1$, or $\mathcal{C}=0$ if $\text{sgn}(\Delta_{\Gamma}\Delta_K)=+1$. To obtain a topologically nontrivial band, the following condition must hold: $|\lambda_2|>|\lambda_{\text{so}}|/2$ (if $\text{sgn}[\lambda_{\text{so}}\lambda_{2}]=1$), or $|\lambda_2|>|\lambda_{\text{so}}|/6$ (if $\text{sgn}[\lambda_{\text{so}}\lambda_{2}]=-1$). Although typically $|\lambda_2|\ll |\lambda_{\text{so}}|$, the latter condition is satisfied for WBr$_3$, which possesses strong SOC. This is in agreement with the Wilson loop calculation shown in Fig.~\ref{fig3}(e).

\section{Hofstadter butterfly of FCB}
\begin{figure}[t]
\begin{center}
\includegraphics[width=3.4in,clip=true]{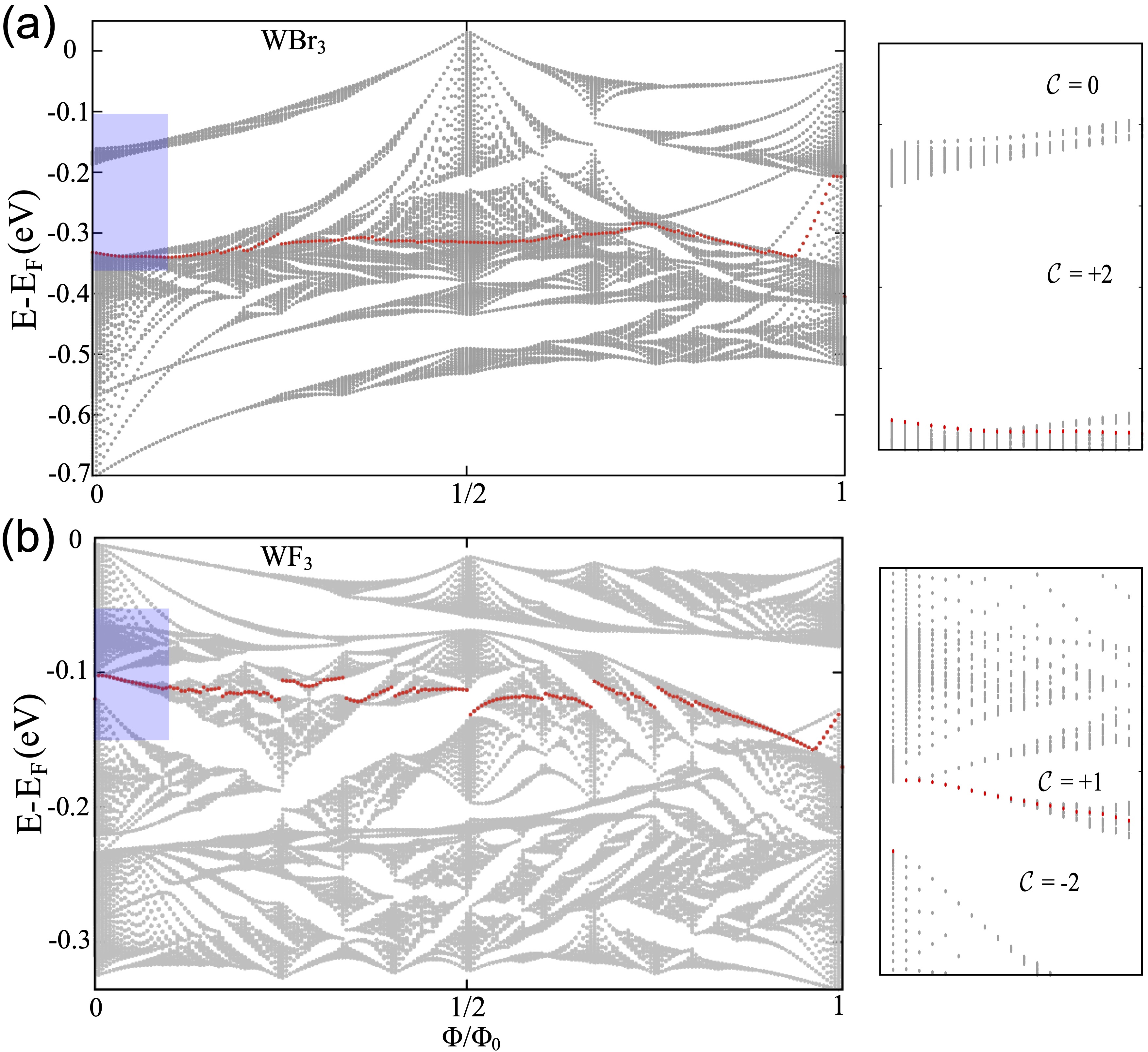}
\end{center}
\caption{Hofstadter butterfly of the three-band model for (a) WBr$_3$ and (b) WF$_3$, where $\Phi/\Phi_0$ is the magnetic flux per unit cell. The $\nu=-1$ filling is labeled by red dots. The blue parts are enlarged on the right panel.}
\label{fig4}
\end{figure}

We now investigate the Hofstadter butterfly and Landau levels of FCB for WBr$_3$ and WF$_3$, as shown in Fig.~\ref{fig4}. Since the FCB is not perfectly flat, and the valence band top in both WBr$_3$ and WF$_3$ exhibits nearly vanishing Berry curvature, the Landau levels at small magnetic flux $\Phi/\Phi_0$ near the Fermi level behave similarly to those in a trivial band structure. Moreover, the Hofstadter spectrum in Fig.~\ref{fig4} reveals an insulator-metal transition at a filling of $\nu=-1$ upon applying a magnetic field. This phenomenon can be understood via Streda formula: $\partial \rho/\partial B=\mathcal{C}/\Phi_0$, which states that the carrier density $\rho$ (or filling) in a state with a fixed, nonzero Chern number $\mathcal{C}$ must change as the magnetic flux $\Phi$ increases. Consequently, at a fixed filling, the Chern insulator gap must exhibit a discontinuity between $\Phi=0$ and finite $\Phi$~\cite{lian2020,herzog-Arbeitman2020}.

\section{Discussions}
The essential ingredient for realizing a FCB with a higher Chern number in these materials is the $sd^2$ hybridization, which transform the atomic $d$ orbitals on hexagonal lattice into $(\tilde{s},\tilde{p}_+,\tilde{p}_-)$ orbitals on triangular lattice. Interestingly, $sd^2$ hybridization on hexagonal lattice can also give rise to an emergent kagome lattice featuring an $\tilde{s}$ orbital, as previously explored in~\cite{miao2014,liu2022}. WF$_3$ and WBr$_3$ serve as two representative cases that exhibit distinct topological mechanisms within $(\tilde{s},\tilde{p}_+,\tilde{p}_-)$ basis. The topological gap in WF$_3$ opens at a finite wave vector $\delta k$ away from band inversion point, which is smaller compared to atomic SOC $E_{g}/E_{\text{inv}}\sim\delta ka$. While the topological gap in WBr$_3$ originates from atomic SOC-induced gap opening at Dirac point. Indeed, the band isolation of FCB from nearby bands in WF$_3$ (with a gap of $\sim36$~meV) is significantly smaller than in WBr$_3$ (where the gap is $\sim143$~meV), further highlighting the distinct topological mechanisms in these materials. Interestingly, both topology and magnetism in these materials emerge purely from $d$ orbitals, distinguishing them from other known topological systems. Given the universal nature of these mechanisms, we anticipate that similar physics should apply broadly to a large class of transition metal trihalides with $P$-$31m$ symmetry.

The interaction energy scale is generated from the lattice constant as $U\sim e^2/\epsilon a$, where $\epsilon$ is the dielectric constant~\cite{koshino2018maximally}. Choosing $\epsilon=10$, we estimate $U\sim0.3$~eV. For isolated FCB in WBr$_3$ and WF$_3$, the bandwidth is significantly smaller than the Coulomb repulsion energy, yielding a ratio of $U/W\gtrsim5$. This suggests that these 2D materials provide a promising platform for exotic correlated states. To evaluate the suitability of FCBs for fractionalized phases at partial filling, two band geometry indicators are employed, namely Berry curvature fluctuation $\delta\mathcal{B}$ and average trace condition violation $\mathrm{T}$ (non-negative) defined as
\begin{eqnarray}
(\delta\mathcal{B})^2 &\equiv& \frac{\Omega_{\text{BZ}}}{4\pi^2}\int_{\text{BZ}} d^2\mathbf{k} \left(\mathcal{B}(k)-\frac{2\pi\mathcal{C}}{\Omega_{\text{BZ}}}\right)^2,
\\
\mathrm{T} &\equiv& \frac{1}{2\pi}\int_{\text{BZ}} d^2\mathbf{k} \left[\mathrm{Tr}(g_{\mu\nu}(\mathbf{k}))-\left|\mathcal{B}(\mathbf{k})\right|\right],
\end{eqnarray}
where $\mathcal{B}(\mathbf{k})\equiv-2\mathrm{Im}(\eta^{xy})$ is Berry curvature, $g_{\mu\nu}(\mathbf{k})\equiv\mathrm{Re}(\eta^{\mu\nu})$ is the Fubini-Study metric, $\eta^{\mu\nu}(\mathbf{k})\equiv\langle \partial^\mu u_{\mathbf{k}}| \left(1- |u_{\mathbf{k}}\rangle\langle  u_{\mathbf{k}}|\right) |\partial^\nu u_{\mathbf{k}}\rangle$ is the quantum geometric tensor. $\mathcal{C}\equiv(1/2\pi)\int d^2\mathbf{k}\mathcal{B}(\mathbf{k})$, $\Omega_{\text{BZ}}$ is area of Brillouin zone. A Chern band with $\mathrm{T}=0$ may be exactly mapped to a  Landau level, which is said to be ``ideal'' for the realization of fractional Chern insulators~\cite{roy2014,claassen2015,ledwith2020,wang2021,parameswaran2013,ledwith2023}. For WF$_3$, $\delta\mathcal{B}=5.38$ and $\mathrm{T}=4.54$. For WBr$_3$, $\delta\mathcal{B}=0.89$ and $\mathrm{T}=0.15$. Notably, the $\mathrm{T}$ value for WBr$_3$ is comparable with those identified in moir\'e materials~\cite{ledwith2020,wang2024,xu2024}. Combined with the higher Chern number, these systems could offer new playground for exotic topologically nontrivial many-body states.

Furthermore, by constructing homobilayers or heterobilayers consisting of WBr$_3$/WF$_3$ or with other 2D semiconductors, one may further tune the bandwidth as well as Chern number of the minibands from isolated Chern band. Also, by stacking WBr$_3$ or WF$_3$ on superconducting 2D materials, such as NbSe$_2$, the chiral topological superconducting phase may be achieved~\cite{wang2015}. We leave all these for future studies.

\begin{acknowledgments}
This work is supported by the Natural Science Foundation of China through Grants No.~12350404 and No.~12174066, the Innovation Program for Quantum Science and Technology through Grant No.~2021ZD0302600, the Science and Technology Commission of Shanghai Municipality under Grants No.~23JC1400600, No.~24LZ1400100 and No.~2019SHZDZX01, and is sponsored by ``Shuguang Program'' supported by Shanghai Education Development Foundation and Shanghai Municipal Education Commission. K.B. and H.W. contributed equally to this work.
\end{acknowledgments}

\end{document}